# Cluster Based Symbolic Representation for Skewed Text Categorization

Lavanya Narayana Raju, Mahamad Suhil, D S Guru and Harsha S Gowda

Department of Studies in Computer Science, University of Mysore, Mysore, India.
`swaralavz@gmail.com, mahamad45@yahoo.co.in, dsg@compsci.uni-mysore.ac.in and harshasgmysore@gmail.com`

**Abstract.** In this work, a problem associated with imbalanced text corpora is addressed. A method of converting an imbalanced text corpus into a balanced one is presented. The presented method employs a clustering algorithm for conversion. Initially to avoid curse of dimensionality, an effective representation scheme based on term class relevancy measure is adapted, which drastically reduces the dimension to the number of classes in the corpus. Subsequently, the samples of larger sized classes are grouped into a number of subclasses of smaller sizes to make the entire corpus balanced. Each subclass is then given a single symbolic vector representation by the use of interval valued features. This symbolic representation in addition to being compact helps in reducing the space requirement and also the classification time. The proposed model has been empirically demonstrated for its superiority on bench marking datasets viz., Reuters 21578 and TDT2. Further, it has been compared against several other existing contemporary models including model based on support vector machine. The comparative analysis indicates that the proposed model - outperforms the other existing models.

## 1 Introduction

With the advancement of digital technology, the amount of text content available on the web has become unimaginably big. Automatic text categorization systems are being developed since last three decades in order to effectively manage such a huge quantity of text documents. Text categorization (TC) is the process of classifying a huge collection of text documents into predefined categories. It carries higher importance due to its huge impact on subsequent activities of text mining and also due to many applications involving text categorization such as spam filtering in emails, classification of medical documents, sentiment analysis etc., (Harish et al., 2010, Aggarwal and Zhai, 2012).

Curse of dimensionality, preserving semantics and effective representation are the issues which make the problem of text classification a suitable one. The complexity of the problem gets doubled if the text corpora are skewed or imbalanced as a class with a large number of samples normally dominates the classes with a small number of samples. In this paper, we address this problem of skewness by transforming imbalanced corpora into balanced corpora through application of a clustering algorithm class wise. Subsequent to partitioning of a large class into a number of smaller sized subclasses, we also recommend, to have a compact representation of text documents by the use of interval valued features. The compact representation not only supports achieving reduction in storage, but also in efficient way of classification through a simple symbolic classifier (Guru and Nagendraswamy., 2006). Nevertheless, to overcome the curse of dimensionality, we adapt a novel representation scheme proposed in (Isa et al., and Guru and Suhil., 2015) which reduces the dimension of the vector space into the number of classes present in the collection.

Overall organization of the paper is as follows. In section 2 we present brief summary of existing works. The proposed model is presented in section 3. The experimental results are discussed in section 4 followed by conclusions in section 5.

## 2 Related Works

From the literature we can understand that, the effort to design systems for automatic text categorization has the history of more than two decades (Hotho et al., 2005; Aggarwal and Zhai, 2012). Machine learning based TC systems carry the following general structure. All the training documents are preprocessed using stemming, pruning, stopwords removal to retain only content terms. Then a matrix representation to the entire training data is given using vector space model which uses the bag of words (terms) (Li and Jain, 1998; Rigutini, 2004). The dimension of such a matrix will be very high even for a dataset of reasonable size which makes the learning algorithms less effective. Hence, dimensionality reduction has been widely explored on text data as a mandatory step in the design of TC to increase the classification performance in addition to reduction in the dimension (Guyon and Elisseeff, 2003).

Most of the works in literature of TC have used either feature selection through ranking or feature extraction through transformation as the means of dimensionality reduction. A number of works can be traced in recent years addressing the problem of text classification through feature selection. Feature selection algorithms such as chi-square, information gain, and mutual information (Yang and Pedersen., 1997) though seem to be powerful techniques for text data, a number of novel feature selection algorithms based on genetic algorithm (Bharti and Singh., 2016; Ghareb et al., 2016), ant colony optimization (Dadaneh et al., 2016; Moradi and Gholampour., 2016; Uysal., 2016; Meena et al., 2012), Bayesian principle (Jiang et al., 2016; Zhang et al., 2016; Feng et al., 2012; Fenga et al., 2015; Sarkar et al., 2014), clustering of features (Bharti and Singh., 2015), global information gain (Shang et al., 2013), adaptive keyword (Tasci and Gungor., 2013), global ranking (Pinheiro et al., 2012; Pinheiro et al., 2015) are proposed.

On the other hand, a classifier is trained and evaluated with a small set of features obtained after dimensionality reduction (Sebastiani., 2002). Thus, it is a very long and time consuming process. However, many applications do not provide such a huge processing capability but still expect the process of classification to be very quick and effective. For such type of applications, it is essential to design a simple yet effective TC system which can predict the probable class of a test document quickly.

It shall be observed from the literature survey, that the existing works are being generic in nature, perform well on balanced text corpora, and are not that effective for skewed / imbalanced text corpora. Hence here in this work, our objective is to convert an imbalanced text corpus into a balanced one through clustering of text documents classwise and then giving it a compact representation.

## 3 Proposed model

The proposed model has two major stages learning and classification.

### 3.1 Learning

The learning stage has 3 different stages; (i) representation of the text documents in lower dimensional space (Isa et al., 2008; Guru and Suhil., 2015) (ii) clustering, where the documents of large sized classes are clustered into sub groups to overcome the problem of skewness, and (iii) compact representation of documents, where a cluster of documents is given a single vector of interval valued feature representation.

#### 3.1.1 Representation of the Text Documents in Lower Dimensional Space

In this section, we present the representation scheme adapted for the text documents. We have intentionally kept this description as a separate section so that the reader should clearly understand that we do not represent the documents using conventional vector space model (VSM) using the bag of words (BoW) constructed for the entire collection. This is due to the fact that, VSM leads to a very high dimensional sparse matrix which is not effective if directly used in computations and hence dimensionality reduction has to be applied through either feature selection or transformation (Sebastiani., 2003). To alleviate this problem, Isa et al., (2008) have proposed an effective text representation scheme which can reduce the dimension of the documents equal to the number of classes at the time of representation itself. In addition to this, (Guru and Suhil., 2015) have proposed a novel term-weighting scheme called 'Term_class Relevance Measure (TCR)' to be used with the representation scheme of Isa et al., (2008) for achieving better performance. Hence, we adapt the representation from Isa et al., (2008) with term weighting scheme of (Guru and Suhil., 2015). A brief overview of the representation and weighting scheme is presented in Fig 1.

Consider a huge collection of text documents say '$N$' in number due to '$K$' different semantically meaningful categories $C_1, C_2,..., C_K$. Each document is first preprocessed using stemming and stop word removal to end up with a collection of content terms. In the representation provided by Isa et al., (2008), initially, a matrix $F$ of size $M \times K$ is created for every document in the collection; where, $M$ is the number of terms assumed to be available in the document as shown in Fig. 1. Then, every entry $F(j,i)$ of the matrix denotes weight of $t_j$ with respect to $C_i$ which is computed using TCR measure as follows (Guru and Suhil., 2015).

| Terms in $d$ | $C_1$ | $C_2$ | ... | $C_K$ |
|---|---|---|---|---|
| $t_1$ | $W(t_1,C_1)$ | $W(t_1,C_1)$ | ... | $W(t_1,C_K)$ |
| $t_2$ | $W(t_2,C_1)$ | $W(t_2,C_2)$ | ... | $W(t_2,C_K)$ |
| ⋮ | ⋮ | ⋮ | ⋮ | ⋮ |
| $t_M$ | $W(t_M,C_1)$ | $W(t_M,C_2)$ | ... | $W(t_M,C_K)$ |
| Feature Vector | $\dfrac{\sum W(t_l,C_1)}{M}$ | $\dfrac{\sum W(t_l,C_2)}{M}$ | ... | $\dfrac{\sum W(t_l,C_K)}{M}$ |

**Fig 1.** Representation scheme for a document $d$

TCR is defined as the ability of a term $t_i$ in classifying a document $D$ as a member of a class $C_j$ as given in (1).

$$TCR(t_i, C_j) = c \times Class\_TermWeight(t_i, C_j) \times Class\_TermDensity(t_i, C_j) \qquad (1)$$

Where $c$ is the proportionality constant defined as the weight of the class $C_j$ as given in (2). Class_TermWeight and Class_TermDensity are respectively the weight and density of $t_j$ with respect to the class $C_j$ which are computed using equation (3) and (4) respectively.

$$ClassWeight(C_j) = \frac{\# Documents\ in\ C_j}{\# Documents\ in\ Training\ Set} \qquad (2)$$

$$Class\_TermWeight(t_i, C_j) = \frac{\# documents\ in\ C_j\ containing\ t_i}{\# documents\ in\ the\ training\ set\ containing\ t_i} \qquad (3)$$

$$Class\_TermDensity(t_i, C_j) = \frac{\# occurences\ of\ t_i\ in\ C_j}{\# occurences\ of\ t_i\ in\ the\ training\ collection} \qquad (4)$$

Then, a feature vector $f$ of dimension $K$ is created as a representative for the document where $f(i)$ is the average of the relevancies of all its terms to $C_i$ from $F$. The main advantage of this representation is that, a document with any number of terms is represented with a feature vector of dimension equal to the number of classes in the population and which is negligibly small in contrast to the feature vector that is created in any other VSM. Therefore, a great amount of dimensionality reduction is achieved at the time of representation itself without the application of any dimensionality reduction technique.

### 3.1.2 Clustering

Availability of balanced text corpora plays a crucial role in the success of any text classification system. This is due to the fact that during the process of training a classifier, the classes with a more number of samples will dominate generally the other classes with a less number of training samples. One solution to handle the class imbalance problem is to convert the corpus into a balanced one. It is true in most of the cases of text classification problems that the variation within a class will increase with the increase in the size of the class. Hence, we perform clustering of documents within each class to convert the class into a collection of dense groups, subclasses. In the process, we also ensure that the sizes of clusters are more or less same.

Given a collection of $N$ labeled documents belonging to $K$ classes say, $C_1, C_2, ..., C_K$ where $i^{th}$ class $C_i$ contains $N_i$ number of documents each is represented by $K$ features as described in section 3.1.1. Let $D_i=\{D_{i1}, D_{i2}, ..., D_{Ni}\}$ be the set of documents of the class $C_i$. A class $C_i$ with $N_i$ number of documents is grouped into $Q_i$ number of dense clusters using hierarchical clustering which is denoted by $Cl^i = \{cl_1^i, cl_2^i, ..., cl_{Q_i}^i\}$. The number of clusters is automatically decided using the inconsistency coefficient. The inconsistency coefficient $i_c$ characterizes each link in a cluster tree by comparing its height with the average height of other links at the same level of the hierarchy. The higher the value of this coefficient, the less similar the objects connected by the link. The value inconsistency coefficient $i_c$ is empirically decided for each class.

Let $Q_1, Q_2, ..., Q_K$ respectively be the number of clusters obtained for the $K$ different classes and let $Q = \sum_{i=1}^{K} Q_i$ be the total number of clusters. For imbalanced datasets the number of clusters obtained will vary from one class to the other class based on the size and variations within a class. Then, a cluster itself can be treated as an independent class and hence the original $K$-class classification problem on an imbalanced corpus has thus become a $Q$-class classification problem on a balanced corpus.

### 3.1.3 Compact Representation

Recently, the concept of symbolic data analysis has gained much attention by the community of researchers since it has proven its effectiveness and simplicity in designing solutions for many pattern recognition problems (Nagendraswamy and Guru., 2007; Punitha and Guru., 2008; Guru and Prakash., 2009;). We can also trace a couple of attempts for text classification using the concepts of symbolic representation and classification (Guru et al., 2010., Harish et al., 2010, Revanasiddappa et al., 2014). In this section, we propose to use interval valued type symbolic data to effectively capture the variations within a cluster of text documents. Another advantage of having such a representation is its simplicity in classifying an unknown document. Given a cluster $cl_j^i$ of a class $C_i$ with $N_j^i$ number of documents $D^{ij} = \{D_1^{ij}, D_2^{ij}, ..., D_{N_j^i}^{ij}\}$ it is represented by an interval valued symbolic representative vector $R_{ij}$ as follows.

Let every document is represented by a feature vector of dimension $K$ given by $\{f_1, f_2, ... f_K\}$. Then, with respect to every feature $f_s$, the documents of the cluster are aggregated in the form of an interval $[\mu^s - \sigma^s, \mu^s + \sigma^s]$ where, $\mu^s$ and $\sigma^s$ are respectively the mean and standard deviation of the values of $t_s$ in the cluster. Hence, $R_{ij}$ contains $K$ intervals corresponding to the K features as,

$$R^{ij} = \{R_1^{ij}, R_2^{ij}, ..., R_K^{ij}\}$$

where, $R_s^{ij} = [\mu^s - \sigma^s, \mu^s + \sigma^s]$ is the interval formed for the $s^{th}$ feature of the $j^{th}$ cluster $cl_j^i$ of the $i^{th}$ class $C_i$. This process of creation of interval valued symbolic representative is applied on all the $Q$ clusters individually to obtain $Q$ interval representative vector $\{R^{11}, R^{12}, ..., R^{1Q_1}, R^{21}, R^{22}, ..., R^{2Q_2}, ..., R^{K1}, R^{K2}, ..., R^{KQ_K}\}$ which are then stored in the knowledgebase for the purpose of classification.

### 3.2 Classification

Given an unlabeled text document $D_q$ its class label is predicted by comparing it with all the representative vectors present in the knowledgebase. Initially, $D_q$ is converted and represented as a feature vector $\{f_1^q, f_2^q, ..., f_K^q\}$ of dimension $K$ as explained in section 3.1.1. Then the similarity between the crisp vector $D_q$ and an interval based representative vector $R$ is computed using the similarity measure proposed by (Guru and Nagendraswamy., 2006) as follows.

$$SIM(D_q, R) = \frac{1}{K} \sum_{s=1}^{K} SIM(D_q^s, R_s)$$

where,

$$SIM(D_q^s, R_s) = \begin{cases} 1 & \text{if } (\mu^s - \sigma^s) \leq f_s^q \leq (\mu^s + \sigma^s) \\ \max\left[\frac{1}{1 + abs((\mu^s - \sigma^s) - f_s^q)}, \frac{1}{1 + abs((\mu^s + \sigma^s) - f_s^q)}\right] & \text{otherwise} \end{cases}$$

Similarly, the similarity of $D_q$ with all the $Q$ representative vectors present in the knowledgebase is computed. The class of the cluster $cl_m$ which gets highest similarity with $D_q$ is decided as the class of $D_q$ as shown in equation (5) below.

$$ClassLabel(D_q) = Class(\underset{i,j}{Arg\max}(SIM(D_q, R^{ij}))) \qquad (5)$$

where, $R^{ij}$ is the representative of the $j^{th}$ cluster of the $i^{th}$ class.

## 4 Experimentation and Results

We have conducted a series of experiments to validate the applicability and efficiency of the proposed model. We have also implemented the Support Vector Machines (SVM) based classification to demonstrate the superiority of the proposed model. The performance of the proposed model has been evaluated using Precision, Recall and F-measure in terms of both micro and macro averaging. The following sections describe details about the skewed datasets considered for experimentation and the results obtained.

### 4.1 Text Corpora

To assess the performance of the proposed model, we have conducted experiments on two commonly used benchmarking skewed text corpora viz., Reuters-21578 and TDT2. The Reuters corpus is a collection of 21578 news articles taken from Reuters newswire (available at http://www.daviddlewis.com/resources/testcollections/ reuters21578/). The total number of topics was 135 where a document may belong to multiple classes. In our experiments we use documents form top 10 categories. There are totally 7285 documents distributed into different classes with a large degree of skew as shown in the Fig. 1. The TDT2 corpus ( Nist Topic Detection and Tracking corpus ) consists of data collected during the first half of 1998 and taken from 6 sources including 2 newswires (APW, NYT), 2 radio programs (VOA, PRI) and 2 television programs (CNN, ABC). It consists of 11201 on-topic documents which are classified into 96 semantic categories. In our experiments we have chosen top 20 categories based on the number of documents to arrive at a subset of 8741 documents distributed with high degree of skew for different classes as shown in Fig. 2. In our experiments, we vary the training set from 10 to 80 percent to verify the performance of the classifier.

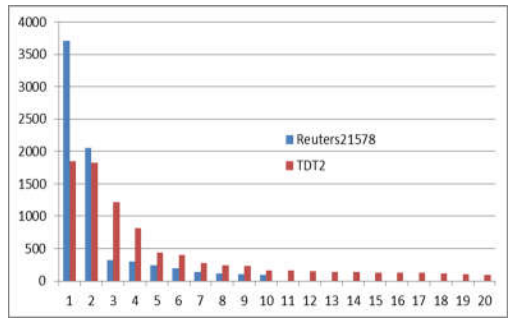

**Fig 2.** Distribution of samples in Reuters-21578 and TDT2 corpora

### 4.2 Results and Discussion

In this section, we present the details of the results obtained by the proposed method and compare it with that of the conventional SVM based classification on Retuers-21578 and TDT2 text corpora. The experiments have been conducted by varying the percentage of training from 10 to 80 percent with 5 random trials each. For every set of training samples, performance is studied in terms of precision, recall and F-measures using both micro and macro averaging.

Fig. 3 and Fig. 4 show the performance of the proposed method in comparison with SVM classifier on Reuters-21578 corpus using macro and micro averaged F measures respectively. It can be observed from the Fig. 3 and Fig. 4 that the proposed method performed well on each training collection than the SVM based classification. Similar study is made on the TDT2 corpus and the results have been shown in Fig. 5 and Fig. 6 for macro averaged and micro averaged F-measures respectively. The similar observation can be made for the TDT2 corpus also as the performance of the proposed method is better when compared to that of the SVM based method.

Further, we have also studied the classwise performance of the proposed method along with SVM classifier based method in terms of precision and recall. This helps in demonstrating the performance of the proposed method with respect to larger and smaller classes and to compare it with that of the SVM classifier. Fig. 7 and Fig. 9 show the performance in terms of precision respectively for SVM classifier and the proposed method, whereas, Fig. 8 and Fig. 10 show the respective performances in terms of recall values on Reuters-21578 corpus. Similarly, Fig. 11 and Fig. 13 show the performance in terms of precision respectively for SVM classifier and the proposed method whereas Fig. 12 and Fig. 14 show the respective performances in terms of recall values on TDT2 corpus. Overall observations made from all the figures showing classwise performances with the increase in the number of features are as follows. For SVM based classification, the value of precision has dropped suddenly for small classes and has reached maximum for large classes whereas, the value of recall has seen an increase for small classes and sudden drop for large classes. On contrary to this, the proposed method has initially seen increase in the performance till 500 features followed by a steady performance thereafter both in terms of precision as well as Recall and hence the overall performance in terms of F-measure is improved significantly.

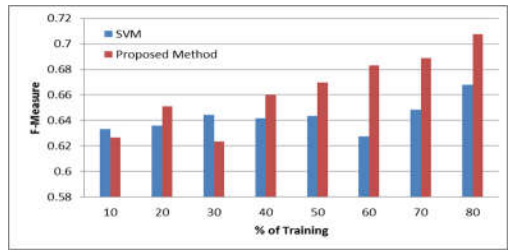

**Fig 3.** Comparison of performances of the proposed model and SVM based model for Reuters-21578 dataset using Macro averaged F-measure

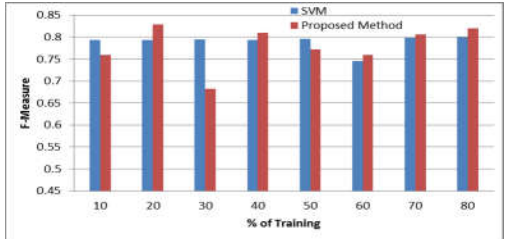

**Fig 4.** Comparison of performances of the proposed model and SVM based model for Reuters-21578 dataset using Micro averaged F-measure

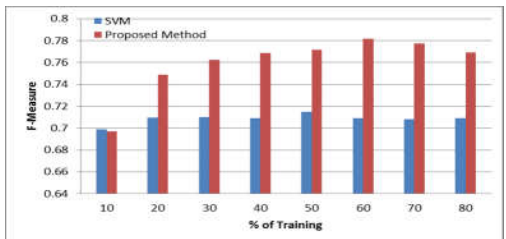

**Fig 5.** Comparison of performances of the proposed model and SVM based model for TDT2 dataset using Macro averaged F-measure

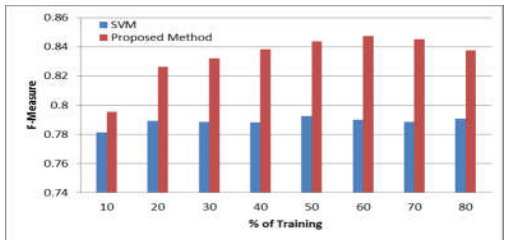

**Fig 6.** Comparison of performances of the proposed model and SVM based model for TDT2 dataset using Micro averaged F-measure

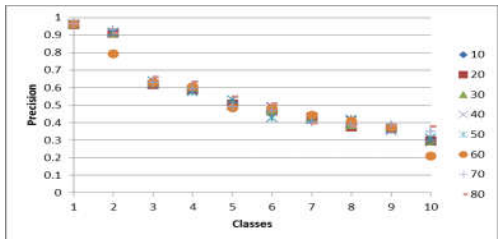

**Fig 7.** Classwise Precision obtained by SVM classifier on Reuters-21578 corpus under varying percentage of training samples

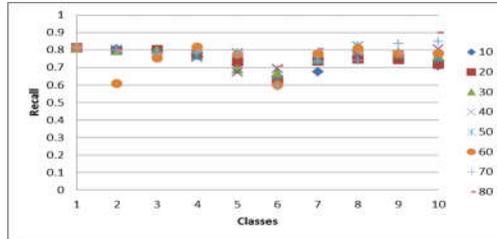

**Fig 8.** Classwise Recall obtained by SVM classifier on Reuters-21578 corpus under varying percentage of training samples

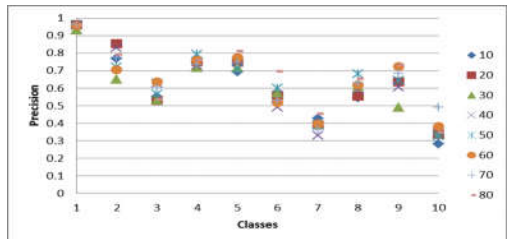

**Fig 9.** Classwise Precision obtained by the proposed model on Reuters-21578 corpus under varying percentage of training samples

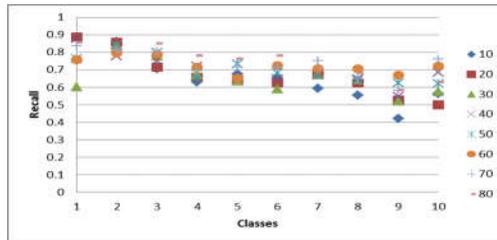

**Fig 10.** Classwise Recall obtained by the proposed model on Reuters-21578 corpus under varying percentage of training samples

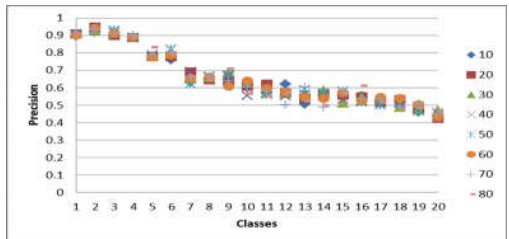

**Fig 11.** Classwise Precision obtained by SVM classifier on TDT2 corpus under varying percentage of training samples

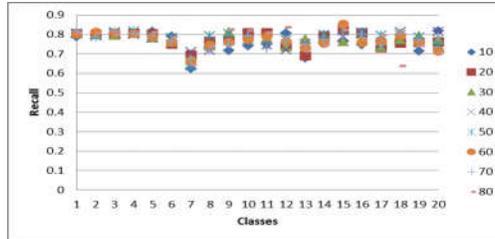

**Fig 12.** Classwise Recall obtained by SVM classifier on TDT2 corpus under varying percentage of training samples

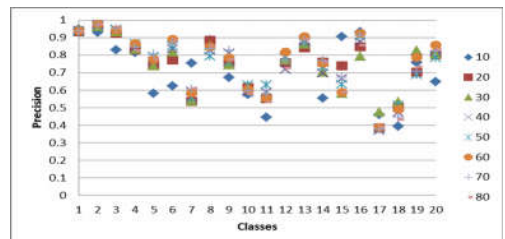

**Fig 13.** Classwise Precision obtained by the proposed model on TDT2 corpus under varying percentage of training samples

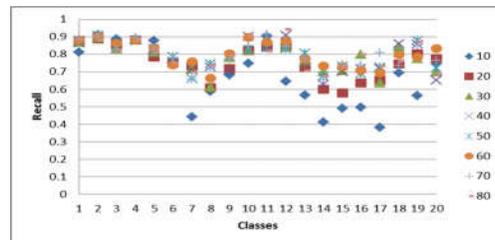

**Fig 14.** Classwise Recall obtained by the proposed model on TDT2 corpus under varying percentage of training samples

Table 1 shows the best performance obtained by the proposed method for both Reuters-21578 and TDT2 corpora in terms of macro and micro F-measures. Also, we have shown the maximum number of clusters formed with respect to each corpus. It can be observed that the number of clusters formed is very less when compared to the total number of training samples considered for training. In Table 2, we also compare the results of the proposed method with that of the state of the art methods. It can be seen from Table 2 that the proposed method outperforms most of the contemporary methods in addition to being very effective since it works with only $K$ features (where $K$ is the number of classes present in the corpus) whereas the other methods need at least few hundreds of features to achieve better performance.

**Table 1:** The best performance of the proposed method on Reuters-21578 and TDT2 corpora in terms of Macro-F and Macro-F

| Text Corpus | No of Training Samples | Maximum No. of Clusters Formed | Macro-F | Micro-F |
|---|---|---|---|---|
| Reuters-21578 | 5828 | 636 | 70.75 | 82.03 |
| TDT2 | 7867 | 582 | 76.92 | 83.74 |

**Table 2:** Comparison of the results of the proposed method with the state of the art techniques on Reuters 21578

| Author and Year | Method | Macro-F (No. of Features) | Micro-F (No of Features) |
|---|---|---|---|
| Uysal., 2016 | IG + IGFSS + SVM | 67.53(500) | 86.473(300) |
| Uysal and Gunal., 2012 | DFS + SVM | 66.55(200) | 86.33(500) |
| Pinheiro et al., (2015) | MFD + BNS | 64(254) | 81.51(254) |
| Pinheiro et al., (2012) | ALOFT + MOR | 62.13 (135) | 80.47 ( 135) |
| Rehman et al., (2015) | DFS, RDC + SVM | 63.47 (500) | 81.98 (500) |
| Aghdam et al., (2009) | ACO | **78.42 ( >= 3600)** | **89.08 ( >= 3600)** |
| **Proposed Method** | **Reduced Representation + clustering + symbolic representation** | **70.75 (10)** | **82.03 (10)** |

## 5 Conclusions

In this work, a method of converting an imbalanced text corpus into a balanced one is presented by exploiting the notion of data clustering. The problem due to skewness of a corpus is addressed. For the purpose of overcoming the curse of dimensionality, we just have adopted our previous model which accomplishes the reduction while representation of documents itself. A method of compact representation of text data by the use of interval-valued data representation is presented in a feature space of dimension equal to the number of classes. It has been experimentally argued that the proposed model is effective in addition to being simple. A comparative analysis indicates that the proposed model outperforms several other existing contemporary models. The finding of this work is that splitting of a larger class of text documents into several smaller subclasses during training would enhance the performance of a classifier.

**Acknowledgements**

The second author of this paper acknowledges the financial support rendered by the University of Mysore under UPE grants for the High Performance Computing laboratory. The first and fourth authors of this paper acknowledge the financial support rendered by Pillar4 Company, Bangalore.